%% file: main.tex
\documentclass[sigconf, nonacm]{acmart}
\def\BibTeX{{\rm B\kern-.05em{\sc i\kern-.025em b}\kern-.08emT\kern-.1667em\lower.7ex\hbox{E}\kern-.125emX}}
    
\usepackage{nicefrac}
\usepackage{siunitx}
\usepackage{array,framed}
\usepackage{booktabs}
\usepackage{
  color,
  float,
  epsfig,
  wrapfig,
  graphics,
  graphicx,
  subcaption
}
\usepackage{textcomp}
\usepackage{setspace}
\usepackage{latexsym,fancyhdr,url}
\usepackage{enumerate}
\usepackage{algorithm2e}
\usepackage{algpseudocode}
\usepackage{graphics}
\usepackage{xparse} 
\usepackage{xspace}
\usepackage{multirow}
\usepackage{csvsimple}
\usepackage{balance}

\usepackage{
  tikz,
  pgfplots,
  pgfplotstable
}
\usepackage{hyperref}

\usetikzlibrary{
  shapes.geometric,
  arrows,
  external,
  pgfplots.groupplots,
  matrix
}

\pgfplotsset{compat=1.9}

\usepackage{mathtools,}

\DeclareMathAlphabet{\mathcal}{OMS}{cmsy}{m}{n}

\DeclareGraphicsExtensions{%
    .png,.PNG,%
    .pdf,.PDF,%
    .jpg,.mps,.jpeg,.jbig2,.jb2,.JPG,.JPEG,.JBIG2,.JB2}

\input{defs}
\begin{document}
\fancyhead{}
\def\thetitle{How BlockChain Can Help Enhance The Security And Privacy in Edge Computing?}
\title{\thetitle}


\author{Jinyue Song}
\email{jysong@ucdavis.edu}
\affiliation{%
  \institution{University of California, Davis}
}

\author{Tianbo Gu}
\email{tbgu@ucdavis.edu}
\affiliation{%
  \institution{University of California, Davis}
}

\author{Prasant Mohapatra}
\email{pmohapatra@ucdavis.edu}
\affiliation{%
  \institution{University of California, Davis}
}

\date{}

\input{abstract}

\maketitle

\input{intro}    
\input{edgeComputing}
\input{blockchain}
\input{blockAndEdge}

\input{futureAndConclusion}

\bibliographystyle{ACM-Reference-Format}
\bibliography{bib}


\end{document}

%% file: defs.tex
\usepackage{xparse}
\newcommand{\bnm}{\begin{newmath}}
\newcommand{\enm}{\end{newmath}}

\newcommand{\bea}{\begin{eqnarray*}}%
\newcommand{\eea}{\end{eqnarray*}}%

\newcommand{\bne}{\begin{newequation}}
\newcommand{\ene}{\end{newequation}}

\newcommand{\bal}{\begin{newalign}}
\newcommand{\eal}{\end{newalign}}

\newenvironment{newalign}{\begin{align}%
\setlength{\abovedisplayskip}{4pt}%
\setlength{\belowdisplayskip}{4pt}%
\setlength{\abovedisplayshortskip}{6pt}%
\setlength{\belowdisplayshortskip}{6pt} }{\end{align}}

\newenvironment{newmath}{\begin{displaymath}%
\setlength{\abovedisplayskip}{4pt}%
\setlength{\belowdisplayskip}{4pt}%
\setlength{\abovedisplayshortskip}{6pt}%
\setlength{\belowdisplayshortskip}{6pt} }{\end{displaymath}}

\newenvironment{newequation}{\begin{equation}%
\setlength{\abovedisplayskip}{4pt}%
\setlength{\belowdisplayskip}{4pt}%
\setlength{\abovedisplayshortskip}{6pt}%
\setlength{\belowdisplayshortskip}{6pt} }{\end{equation}}

\newcounter{ctr}

\newcounter{mytable}
\def\mytable{\begin{centering}\refstepcounter{mytable}}
\def\endmytable{\end{centering}}

\newcounter{myfig}
\def\myfig{\begin{centering}\refstepcounter{myfig}}
\def\endmyfig{\end{centering}}

\newlength{\saveparindent}
\newlength{\saveparskip}
\newcommand{\E}{{\rm I\kern-.3em E}}

\renewcommand{\eqref}[1]{\mbox{Equation~(\ref{#1})}}

\def \part {part}

\renewcommand{\paragraph}[1]{\vspace*{6pt}\noindent\textbf{#1}\;}

\def \blackslug{\hbox{\hskip 1pt \vrule width 4pt height 8pt
    depth 1.5pt \hskip 1pt}}
\def \qed{\quad\blackslug\lower 8.5pt\null\par}

\newcounter{mynote}[section]

\newcommand\ignore[1]{}

\newcounter{rcnote}[section]

\newcounter{mrnote}[section]

\newcounter{fknote}[section]

\newcounter{anote}[section]

\DeclareMathSymbol{\mlq}{\mathord}{operators}{``}
\DeclareMathSymbol{\mrq}{\mathord}{operators}{`'}

\newcommand{\rhf}[2]{R_{f, \gamma}}

\DeclareDocumentCommand{\edist}{o o}{
  \ensuremath{
    \IfNoValueTF{#1}{{d}}{{\sf d}(#1,#2)}
  }
}

\newcommand{\olrk}[1]{\ifx\nursymbol#1\else\!\!\mskip4.5mu plus 0.5mu\left(\mskip0.5mu plus0.5mu #1\mskip1.5mu plus0.5mu \right)\fi}

\NewDocumentCommand{\indseq}{ O{1} O{r} }{{#1}\ldots {#2}}


%% file: abstract.tex
\begin{abstract}
In order to solve security and privacy issues of centralized cloud services, the edge computing network is introduced, where computing and storage resources are distributed to the edge of the network. However, native edge computing is subject to the limited performance of edge devices, which causes challenges in data authorization, data encryption, user privacy, and other fields. Blockchain is currently the hottest technology for distributed networks. It solves the consistent issue of distributed data and is used in many areas, such as cryptocurrency, smart grid, and the Internet of Things. 

Our work discussed the security and privacy challenges of edge computing networks. From the perspectives of data authorization, encryption, and user privacy, we analyze the solutions brought by blockchain technology to edge computing networks. In this work, we deeply present the benefits from the integration of the edge computing network and blockchain technology, which effectively controls the data authorization and data encryption of the edge network and enhances the architecture's scalability under the premise of ensuring security and privacy.
Finally, we investigate challenges on storage, workload, and latency for future research in this field.

\end{abstract}

%% file: intro.tex
\section{Introduction}
\label{sec:intro}


With the development and popularization of cloud services and smart devices, edge computing\cite{shi2016edge} has penetrated all levels and aspects of our lives, improving our lives and work efficiency. Combined with blockchain technology, the security and privacy of edge computing raise our attention to new research directions.

There are various smart devices in the living and industrial environments connecting to servers without manual intervention, such as medical facilities, smart grid, smart home systems, vehicles and road side unit networks etc. Cloud computing's efficiency cannot support the massive amounts of data\cite{alzahrani2014mobile} generated by these devices and cannot meet the rapid response requirements in certain scenarios. Therefore, edge computing with low latency and heterogeneous is the trend of remote data services development. With service providers' deployment close to the end devices, the request latency may be lower, and the throughput may be higher. Also, using an edge computing network can avoid the potential security and privacy risk in a centralized system. For example, the data processing could happen in the local server instead of the centralized one, lowering the security risk of data leaking. 

However, edge computing is not a perfect design, and it has disadvantages in security and privacy aspects. Due to its heterogeneous and distributed structure, it is challenging to manage various devices and servers in the network. Also, these servers have limited hardware resources which is hard to handle heavy computational tasks.  
It is essential to integrate blockchain technology into edge computing. Blockchain\cite{gupta2017blockchain} is a distributed database design supported by consensus protocols and within this environment, smart contract is a distributed program deployed on each server. Such a design can improve the security and privacy of the system under the premise of ensuring the high efficiency and availability of computing service.

Combined with blockchain technology, edge computing has a good research vision, but there are still many challenges to be solved in this field. Due to edge devices' limited computing and storage resources, ensuring the integrity and validity of blockchain data is a considerable challenge. In addition, due to the heterogeneous organizational form of edge computing, potential security issues may arise, such as the authentication of the local network and the encryption requirements of local user data. These challenges are worthy of our in-depth discussion and research.



%% file: edgeComputing.tex
\section{Edge Computing}\label{edgeComput}

Edge computing refers to computing services provided for end-user devices at the edge of the network, which is the middle layer between cloud services and local devices. It provides fast-response services such as computing, storage, communication, and control in a location close to the data source of the end-user devices.

\subsection{Research Areas in Edge Computing}

We combine multiple study cases with various fields to present the benefits of edge networks.

\subsubsection{Real-Time Online Data Processing}

Compared with traditional centralized cloud services, edge computing has the advantages of fast response and low latency, especially for real-time data processing, which requires very high bandwidth. Real-time data processing services based on the traditional centralized cloud are often saturated with uplink capacity. They cannot satisfy the transmission of large chunks of data, such as high-definition video, which affects the performance of the analysis service. Wang\cite{wang2018bandwidth} proposed a bandwidth-efficient real-time video analysis mechanism for drones, which filters frames, adjusts learning models, and reduces bandwidth usage. Zhou\cite{zhao2012game} provided a locally virtualized CPU/GPU cluster for high-quality graphics in a real-time online game, which has 14 to 38\% better performance in the evaluation.

\subsubsection{Internet of Things}

Combined with edge computing, IoT frameworks can meet the needs of fast deployment and high-speed response and solve the problems of limited storage space and insufficient computing power of IoT devices. Compared with centralized cloud services, edge computing can efficiently process data generated by IoT devices promptly while avoiding the large amount of data transmission required by cloud services.

Chen\cite{chen2018edge} proposed an industrial-level IoT manufacturing architecture based on edge computing. This architecture uses the time-sensitive network (TSN) protocol to manage time-sensitive task data at the network level. Also, sensing devices are close to the factory infrastructure, and the computing and storage nodes in the edge network are used to dynamically adjust the execution strategy of the factory facilities according to the sensor status.

\subsubsection{Vehicular Networks}
When the vehicular network uses cloud services to perform computing tasks between vehicles, there will be problems with network latency and high transmission overhead. Especially when vehicular network data has an explosive growth, cloud services cannot achieve real-time data analysis and feedback, which brings potential safety problems to vehicles in motion. Edge computing provides an alternative solution to deploy computing and storage services near the vehicles\cite{zhang2017mobile}, reducing latency and saving energy consumption.

\subsubsection{Computing Node and Task Management}\label{container}

Scalability is one of the advantages of edge computing networks. Traditional cloud computing has a centralized management node responsible for the data and task scheduling of all devices in the network. When the size of the cloud network scales up, this centralized management node will become the bottleneck of system performance. The cloudlets\cite{ahmed2017bringing} design of edge computing leaves local data for local processing, which cuts down the cloud's bandwidth usage by three to six orders of magnitude while also providing low-latency and high-bandwidth services. This design can introduce more edge nodes while the marginal cost increase is minor, ensuring the system's scalability.

Within one cloudlet, the performance of devices varies greatly, and each cloudlet has different operating status and conditions. A virtualized environment\cite{aruna2020performance}, such as a container, can be used for these various cloudlets and manage in a unified manner. The optimized algorithm\cite{krylovskiy2015internet} enables the edge network to expand its structure while ensuring stable device management.




\subsection{Security and Privacy in Edge Computing}\label{secAndPriInEdge}

\subsubsection{Benefits:}
User privacy and data security are very worthy of attention. Centralized cloud services have brought growing concerns about this field. When raw data is sent to the cloud, it is out of the control of users and organizations. For example, the security camera contains the user's facial information, the online shopping record reflects the user's financial status, and the factory's electricity consumption represents its production status. The centralized server cannot guarantee the whole data security when it is attacked by hackers or misused by third parties.

Edge computing provides an alternative to protecting data\cite{zhang2018data}. The data from the edge device can be encrypted first, where the representative encryption algorithms include: AES\cite{osvik2010fast}, RSA\cite{mahajan2013study}, and other encryption approaches\cite{baharon2015new}. For distributed edge networks, identity authentication\cite{liu2014shared} is another method to protect data, regulating access rights and avoiding illegal access to sensitive data.

\subsubsection{Challenges:}
But relying only on the framework of edge computing, many security and privacy challenges still need to be studied and resolved in depth. First, many IoT devices and sensors in the edge network may have low accuracy, so abnormal and malicious data collected is a significant challenge. Then, because of the heterogeneous design of the edge network, the computing nodes and devices are very diversified, resulting in a coarse degree of data access management, and malicious access may penetrate the core of the edge device. Unverified computing nodes joining the network may hack user data and disrupt the operation of the edge network. In addition, due to the performance limitations of edge nodes, these devices cannot resist network attacks, such as man-in-the-middle attacks and denial of service attacks, which leads to the paralysis of the edge network and loss of stability. 

Attacks and issues that threaten privacy and security often occur in three aspects of edge networks: aiming specific data nodes, aiming data transmission, and model synchronization among nodes and edge clouds. For these three aspects, we will analyze the solutions with representative cases in section \ref{blockchainBasedEdgeComputing}.

%% file: blockchain.tex
\section{Blockchain}\label{blockc}

Blockchain is a fiery research and application field. It is an encrypted distributed anonymous ledger system\cite{gupta2017blockchain} recording transactions between participants, which solves inconsistent information consensus in a distributed network, such as the Byzantine Generals problem. Based on the blockchain platform, smart contracts have realized decentralization, automatic execution, and other functions, enriching the research and application of the blockchain field—for example, cryptocurrency combined with blockchain technology, Bitcoin, and Ethereum.


\subsection{What Security and Privacy Benefits introduced by Blockchain?}

Blockchain technology builds a credible interaction method between untrusted computing nodes, which fits in the distributed structure of edge computing. It removes the dependence on centralized management nodes and supports data privacy and security through encryption and consistency mechanisms such as Proof-of-Work.

The framework of an edge network generally has three layers: the bottom layer is composed of small sensors and devices, the middle layer contains servers with limited data processing and storage capabilities close to the data source, also known as cloudlets, and the upper layer is composed of large data centers. Due to the high computing cost of the blockchain consensus algorithms and the limited performance of the small underlying devices, the blockchain is generally deployed among multiple cloudlets in the middle layer to form a consensus and ensure data security.

About the security and privacy issues mentioned in section \ref{secAndPriInEdge}, we show how blockchain technology can resolve these threats in edge computing networks. When a single edge device or computing node is hacked to broadcast malicious data, these polluted data will not be transmitted through the network since most nodes and devices use the blockchain's distributed ledger to verify and identify the malicious data. Similarly, the read-write permissions can be recorded on the blockchain blocks without a centralized management to verify access permissions, so users and local nodes can have the same knowledge of the global data access from each server. The program operation deployed among nodes can be automatically executed through the decentralized smart contracts, without a central node to trigger.

\subsection{Research Areas Benefiting from Blockchain in Security and Privayc Perspective.}


In this section, we summarize some areas where blockchain technology is utilized and combine study cases to analyze the improvements that blockchain technology brings to edge computing.

\subsubsection{Internet of Things}
Due to the lack of trust between IoT devices and computing nodes, building a trusted relationship within this network is challenging. The distributed ledger of the blockchain is a feasible solution. For example, the device's identifier and mac address will reveal its model number and location\cite{hamdaoui2019unleashing}. This sensitive information can be encrypted and recorded in the immutable block; it protects the device's privacy, and at the same time, it allows other peers to verify the device's identity through the encrypted data extracted from the block, which maintains a trusted communication relationship in an untrusted network.

On the other hand, a decentralized and automatically executed smart contract can process the IoT's program operations without support from a centralized control node\cite{song2020smart}. Then, the output generated by the smart contract will be stored in the block. This design ensures consistent program operations and outputs, avoiding suspicious execution and providing data security.

\subsubsection{Vehicular Cloudlet.}
The biggest difference between vehicles and other types of Internet of Things lies in vehicles' high mobility and Spatio-temporal sensitivity. It is very challenging to maintain the data integrity and accuracy of moving vehicles. With the help of the roadside unit, vehicles can quickly join the network, perform encrypted communications, and synchronize valid vehicle information\cite{liu2018blockchain} with the blockchain consensus mechanism. This procedure ensures that vehicles can maintain data consistency, avoid data loss, and ensure data integrity and validity.

\subsubsection{Payment and Loan.}
Digital encryption currency is the most significant application of blockchain technology. However, the mining work and transactions of digital currencies are processed on traditional networks and servers. The introduction of edge computing frameworks increases the mobility and economical property of the system.

The deployed smart contract can automatically execute computing resource transactions and currency lending\cite{song2020smart}. Because of its transparency in data and execution, developers can analyze the contract's operating status and financial costs. In addition to combining game theory and Nash equilibrium theory\cite{zhang2019joint}, the system can make optimal decisions with minimum costs.

\subsubsection{Privacy-preserved Tracking.}
The tracking function can describe the user's movement trajectory, allowing the system to learn more data and accurately predict the overall users' movement trend. But directly uploading the user's raw data to the system will expose privacy, so additional mechanisms are demanded to reduce the data's sensitivity while maintaining its accuracy.

The anonymity of users in the blockchain environment is the key to protect user's privacy and the decentralized and automatic execution smart contract is the core to avoid malicious operations. Using blockchain technology for traffic tracking\cite{wang2020trafficchain} can protect the location privacy of vehicles while collecting travelling history on a decentralized transportation network. The recent coronavirus infection tracking is also a hot field. Unlike the centralized management of the user's infection status and travel path, the blockchain-based decentralized system\cite{song2020blockchain} can ensure users' anonymity by randomization mechanism, and its smart contract guarantees the virus transmission path data is accurate and secure saved in the block. 




%% file: blockAndEdge.tex
\section{Security and Privacy Enhancement in Edge Computing via introducing Blockchain.}\label{blockchainBasedEdgeComputing}

\begin{figure}
\centering
  \includegraphics[width=\columnwidth]{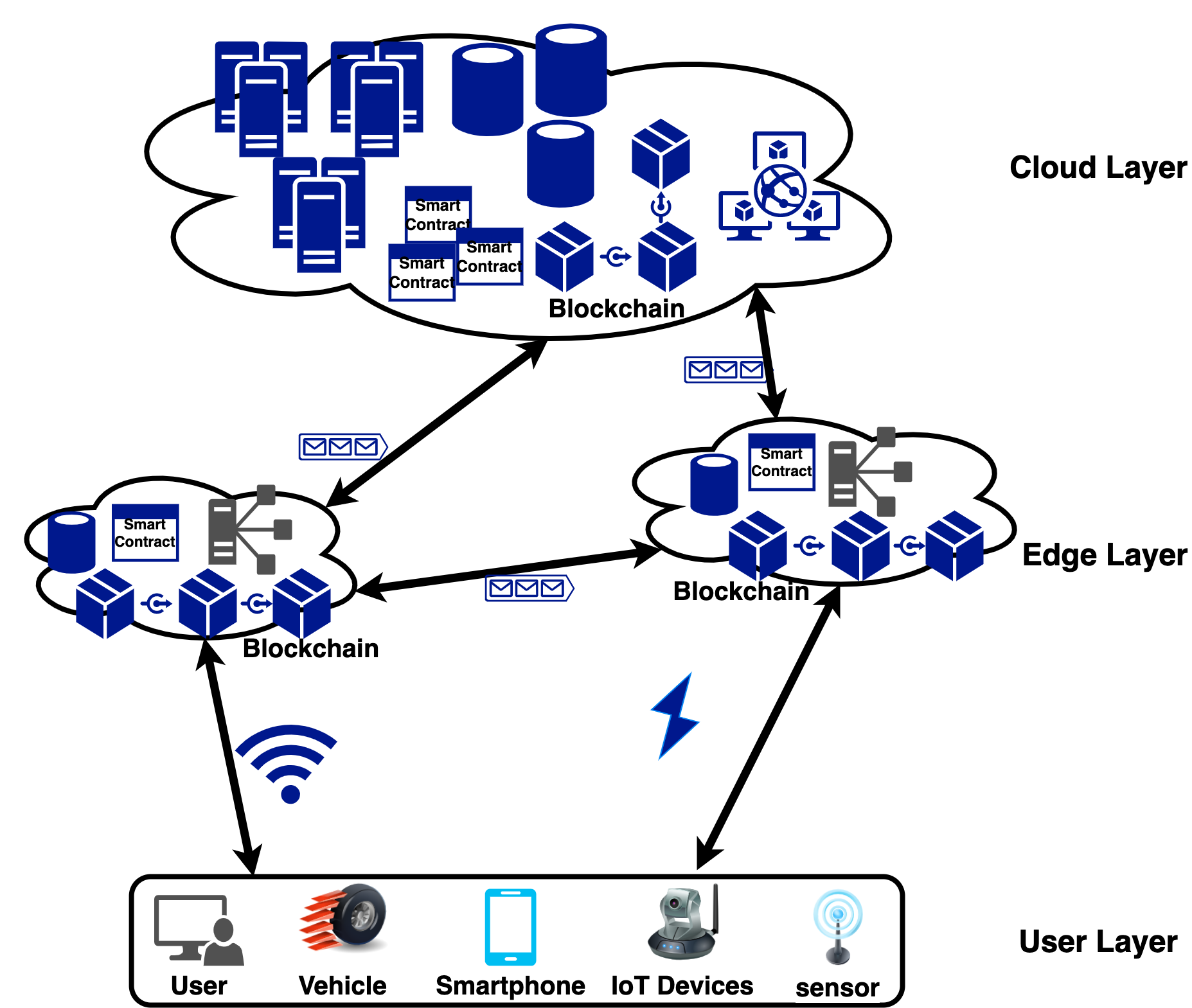}
  \caption{General blockchain based edge computing framework.}~\label{fig:bc_edge}
  \vspace{-0.5 cm}
\end{figure}

Figure \ref{fig:bc_edge} shows a general structure for the edge computing framework based on blockchain technology, which may have three layers: the user layer at the bottom comprises the edge devices as the data source, and the edge layer in the middle is the edge node with limited computing and storage capabilities. In most designs, they undertake the work of blockchain miners and ensure the synchronization of blocks and the execution of smart contracts, and the cloud layer on the top is a large data center that provides cloud services.

In general, a blockchain-based edge computing system has the following working procedure. Firstly, the bottom edge devices provide activity data, including the interactions between peers and between edge nodes. Then, the middle edge node verifies these data and writes them into blocks based on the blockchain consensus mechanism. After that, the large-scale data center on the top layer does the necessary structural analysis and management.
We will analyze the feasibility and benefits of the blockchain-based edge computing system from the security and privacy aspects.

\subsection{Anonymity}
In an untrusted distributed edge network environment, anonymity is an essential approach to protecting the identities of devices and users. The device can rely on smart contracts to verify its identity while using the edge node server as an information transmission intermediary without exposing its identity to the server. This general design is vital for stable and secure communication between the devices and regional cloudlets in the edge network.

In the edge network, the data of the underlying device is provided to the upper nodes. The underlying device's identity information should not be exposed to other insecure networks. There is a framework\cite{usman2019distributed} to centralize and register these underlying devices in the first-level edge device, then it manages the communication and computation among these underlying devices. However, such a scheme makes the level one edge device the local registration collector in the edge network. Once the collector leaks these identities, the anonymity of the locally registered devices will not exist.

Blockchain can provide anonymity to the system and hide the identity of devices in the edge network. Wang\cite{wang2019blockchain} proposed an anonymous identity verification protocol based on the blockchain. The edge unit is registered and verified with the registration authority to ensure that the edge units' identity information will not be exposed in the untrusted network.

\subsection{Authentication}
In the edge network environment, the data generated by devices will interact across regions, which inevitably needs to verify the identity of the devices and manage their data read and write permissions. These records are written on the blockchain, and the smart contract executes the device's identity verification to ensure the security of the data in the edge network.

In the protocol designed by Wang\cite{wang2019blockchain}, the registration and data authorization of the edge unit depend on the execution of the smart contract. Only the registration authority can associate the real identity of the device with the public key, so the authorization and communication between the devices are guaranteed secure.

The dynamic domain name resolution strategy proposed by Guo\cite{guo2019blockchain} provides services for edge authorization in the system. The edge terminal's domain name-related information is stored in the blockchain, and the communication between the terminals is supported by an asymmetric encryption algorithm based on elliptic curve cryptography. Similarly, the lightweight blockchain-based identity verification protocol proposed by Jangirala\cite{jangirala2019designing} allows devices to authenticate each other and establish a session key, which technically guarantees the anonymity of user identities.

\subsection{Mobile Edge Computing}
Mobile edge computing is a novel research field for security and privacy. Edge computing nodes must balance the performance of offloaded computing tasks and the security of local data processing. Excessive data processing requests or other attacks should not cause incomplete data transmission from the mobile edge node, thereby losing data security and user privacy protection.

Xu\cite{xu2019become} combines blockchain technology to provide a feasible resource allocation scheme for computing offloading, which optimizes energy consumption, task offloading time and maintains load balance while ensuring the data integrity and security of edge computing offloading. 
Similarly, Song\cite{song2020blockchain} provides a COVID-19 infection tracing framework based on the smart contracts to monitor the location's infection status. In this design, the smart contract group balances the workload between local and upper-level contracts for user check-in requests, reducing network traffic. 

\subsection{Protocol Security}
The blockchain-based protocol design involves more complex and practical security issues. A protocol design cannot meet all scenarios and requirements. For example, it is challenging to guarantee the untraceability of identity in a protocol while it is lightweight. Therefore, we will analyze the security solutions of the protocol in combination with specific scenarios.

In the IoT supply chain scenario based on mobile edge computing, Jangirala\cite{jangirala2019designing} proposed a lightweight radio frequency identification-based authentication protocol with blockchain support to provide enough bandwidth for real-time data of goods. This protocol is simple to operate, has provable security, and balances communication and computing costs. NKN Lab\cite{nknwebsite} proposes a more general-purpose self-evolving and self-motivating decentralized network, not limited to specific application scenarios. The blockchain in this network has novelly presented Proof of Relay instead of Proof of Work. While reducing energy consumption and improving performance, this network can also resist various attacks, such as Double spending attck and Denial-of- Service attacks, so security is guaranteed in this network.

\subsection{Security and Privacy in Architecture}
In addition to the protocol perspective or implementation of specific applications discussed above, we also need to show the security and privacy contribution of blockchain in the edge structure from the perspective of architectural design.

In Rathore's decentralized network security architecture\cite{rathore2019blockseciotnet}, blockchain and mobile edge networks are its primary technical support. Blockchain can detect decentralized attacks, avoiding the collapse of the architecture due to the crash of one certain node. In addition, the introduction of edge architecture removes storage constraints, reduces computing costs and latency.

The interaction between the layers of the system is also related to security and privacy. Ines proposed a modular and hierarchical design in a complex Internet of Things environment, which generates good positive feedback in practical applications and ensures the security of data storage and computing. Ines' global edge computing architecture\cite{sitton2019review} consists of three layers: IoT layer, Edge layer, and Business solution layer. The transmission data from the IoT layer is processed by encrypted hardware and communication protocol before it is connected to the Edge layer, an open layer containing many different service providers. This Edge layer pre-processes the raw data from IoT devices and transfers it to the Business solution layer. The use of blockchain ensures the data integrity transmitted through layers. At the business solution layer, the system verifies IoT devices' identity by smart contracts and extracts data from the blockchain database to establish a secure computing channel. Therefore, the three levels of security and privacy are guaranteed under the support of hardware and software protocols.


\subsection{Data Security}
In this subsection, we present the advantages of blockchain-based edge networks from the perspective of the definition of data security and its case studies.
The security of data is reflected in three aspects: confidentiality, integrity, and availability. In our scenario, these three aspects have more detailed definitions.
\begin{itemize}
    \item Confidentiality: The data stored at the edge node must be encrypted and protected, and only users with the key can access the data.
    \item Integrity: In a distributed blockchain-based system, the consensus mechanism of the edge node data needs to resist attacks, such as 51\% attacks and Sybil attacks. Data saved on nodes needs to be updated to the latest version within an acceptable time to ensure data consistency.
    \item Availability: Edge nodes should provide data access for edge devices and top layer cloud services at any time. Even if some nodes are shut down, the data can still be recovered on new edge nodes and provide access services.
\end{itemize}

\subsubsection{Confidentiality.} In addition to standard data encryption, we also need to consider the management of data read and write permissions and user identity authentication because illegal reads and writes are also a significant cause of data pollution.

Users need an efficient approach to imply identity authentication to access data in edge node. Ren\cite{ren2020novel} proposed an identity authentication scheme for the trading system, where users' identity information is stored in the block for data consistency and not tamper. The trusted authentication method proposed by Shaoyong\cite{guo2019blockchain} is based on the name resolution and authentication of edge computing nodes, and a caching strategy is designed to improve the performance of the system's authentication mechanism.

\subsubsection{Integrity.} The traditional consensus mechanism is proof-of-work or proof-of-stake, which occupies a lot of computing resources.  It is not practical and cost-effective to imply on edge devices and nodes with limited hardware performance. Therefore, the reputation of devices is an alternative method to determine if the device data be written into the block. For example, the vehicular edge network based on the blockchain\cite{kang2018blockchain} is supported by reputation data sharing scheme to ensure high-quality data sharing between roadside units and vehicles. This design provided a three-weight subjective logic model to accurately manage vehicle reputation and solve the security problems of untrusted participants in the network and malicious data injection.

\subsubsection{Availability.} 

The heterogeneity of edge computing networks has profoundly affected the availability of data. It can unify the communication at different levels of the edge network through software support and hardware virtualization methods. Huan\cite{zhou2021building} proposed the ALLSTAR ecosystem, which contains four subsystems to accomplish this task. Heterogeneity-AS-code toolKIT (HASKIT) and CRoss-Edge seamless infrAstructure orchestration Framework (CREAF) focus on software perspective to establish a model to describe heterogeneous services and provide the relationship between levels Interface. HardWare chAracterized Function vIRtualization framework (WAFIR) focuses on the hardware perspective to unify the virtual environment of hardware. Service Ederation defined Trustworthy Inter-Chain platform (SETIC) is a blockchain-based subsystem to record user transactions and uses smart contracts to execute decentralized data management tasks automatically.

\subsection{User Privacy}

The definition of user privacy in this work is: in the edge computing environment, data mining and cyber-attacks will not expose the user's identity and sensitive information\cite{xu2014information}. Edge devices should control what data is shared at edge nodes, how to verify the validity of the data under the premise of encryption, and who else the data is shared with. We discuss the user privacy protection from the blockchain-based perspective of data encryption methods.

The native cryptography in the blockchain provides a digital signature for users to prove and verify the encrypted data without revealing the plain content, and Elliptic Curve Digital Signature Algorithm (ECDSA)\cite{johnson2001elliptic} is the typical one used on both Bitcoin and Ethereum platform. These native encryption mechanisms can meet the needs of general computing applications, but for applications such as industrial IoT sensing and outdoor drone data synchronization, lightweight and customized protocols are required to meet new scenarios requirements such as reducing computing costs for resource-limited IoT devices.



Except for those traditional encryption mechanisms to protect sensitive and private data, users can encrypt sensitive data and identities in the blockchain transaction system, achieving data security while providing privacy services. We consider removing identifiers and trajectory data is also an encryption approach to protect user privacy. Before publishing the source data, identifiers and sensitive data that can infer users' identities should be removed or added with noise. Nowadays, social media are very popular, and users activities are logged on those platforms and analyzed by third parties. Even users' true identities are hidden, and they can be tagged with particular id numbers for analysis. Each user's activity has a unique pattern to be exposed to even though their true identities are hidden. Another potential challenge comes with this encryption approach. If source data is strongly anonymous, its usability will be lower, and the information is lost. Blockchain provides a balanced option on the encryption of sensitive data and its usability.   
Keke\cite{gai2019permissioned} designed the smart grid's operating architecture. He mapped the grid's network nodes to the blockchain and edge computing network and used pseudo-names to mark users to avoid revealing their real identities.

%% file: futureAndConclusion.tex
\section{Challenges and Future research Directions.}\label{future}


Although blockchain technology provides solutions to the challenges of edge computing from many aspects, the combination of the two fields still has potential issues and new research areas for us to explore.


The first problem is about the storage consumption. It increases a lot because of the blockchain consensus mechanism, which requires a complete Blockchain's distributed ledger data in edge nodes within the network. It is potentially considered a huge storage waste for duplication. 

The second problem is about the unbalanced workload. A powerful computing node sustains the significant mining and computing load of the blockchain, which may not be fair for the node because the digital currency or incentive from private blockchain is not tradable for public. This challenge requires a fair and dynamic resource management among edge nodes. 

The last problem is about the transaction latency. The blockchain requires block synchronization among edge nodes, where they have a competition on writing transactions into the next block. The new block generation time cost delays the system operation. Spatio-temporal sensitive applications may suffer from this kind of latency, for example, moving self-driving vehicles cannot wait seconds for the synchronization of traffic status. 


\section{Conclusion}\label{conclus}
This survey focuses on the security and privacy aspects of blockchain-based edge computing networks. It demonstrates the feasibility of deployment in the Internet of Things, vehicular network, payment, tracking, etc. In particular, we discussed the challenges and benefits of these sub-areas of security and privacy. No universal system can solve all security and privacy issues, and we can only use a more suitable system framework for specific demands.

%% file: main.bbl

\begin{thebibliography}{00}


\ifx \showCODEN    \undefined \def \showCODEN     #1{\unskip}     \fi
\ifx \showDOI      \undefined \def \showDOI       #1{#1}\fi
\ifx \showISBNx    \undefined \def \showISBNx     #1{\unskip}     \fi
\ifx \showISBNxiii \undefined \def \showISBNxiii  #1{\unskip}     \fi
\ifx \showISSN     \undefined \def \showISSN      #1{\unskip}     \fi
\ifx \showLCCN     \undefined \def \showLCCN      #1{\unskip}     \fi
\ifx \shownote     \undefined \def \shownote      #1{#1}          \fi
\ifx \showarticletitle \undefined \def \showarticletitle #1{#1}   \fi
\ifx \showURL      \undefined \def \showURL       {\relax}        \fi
\providecommand\bibfield[2]{#2}
\providecommand\bibinfo[2]{#2}
\providecommand\natexlab[1]{#1}
\providecommand\showeprint[2][]{arXiv:#2}

\bibitem[\protect\citeauthoryear{Ahmed, Ahmed, Yaqoob, Shuja, Gani, Imran, and
  Shoaib}{Ahmed et~al\mbox{.}}{2017}]%
        {ahmed2017bringing}
\bibfield{author}{\bibinfo{person}{Ejaz Ahmed}, \bibinfo{person}{Arif Ahmed},
  \bibinfo{person}{Ibrar Yaqoob}, \bibinfo{person}{Junaid Shuja},
  \bibinfo{person}{Abdullah Gani}, \bibinfo{person}{Muhammad Imran}, {and}
  \bibinfo{person}{Muhammad Shoaib}.} \bibinfo{year}{2017}\natexlab{}.
\newblock \showarticletitle{Bringing computation closer toward the user
  network: Is edge computing the solution?}
\newblock \bibinfo{journal}{{\em IEEE Communications Magazine\/}}
  \bibinfo{volume}{55}, \bibinfo{number}{11} (\bibinfo{year}{2017}),
  \bibinfo{pages}{138--144}.
\newblock


\bibitem[\protect\citeauthoryear{Alzahrani, Alalwan, and Sarrab}{Alzahrani
  et~al\mbox{.}}{2014}]%
        {alzahrani2014mobile}
\bibfield{author}{\bibinfo{person}{Ahmed Alzahrani}, \bibinfo{person}{Nasser
  Alalwan}, {and} \bibinfo{person}{Mohamed Sarrab}.}
  \bibinfo{year}{2014}\natexlab{}.
\newblock \showarticletitle{Mobile cloud computing: advantage, disadvantage and
  open challenge}. In \bibinfo{booktitle}{{\em Proceedings of the 7th Euro
  American Conference on Telematics and Information Systems}}.
  \bibinfo{pages}{1--4}.
\newblock


\bibitem[\protect\citeauthoryear{Aruna and Pradeep}{Aruna and Pradeep}{2020}]%
        {aruna2020performance}
\bibfield{author}{\bibinfo{person}{K Aruna} {and} \bibinfo{person}{G Pradeep}.}
  \bibinfo{year}{2020}\natexlab{}.
\newblock \showarticletitle{Performance and scalability improvement using
  IoT-based edge computing container technologies}.
\newblock \bibinfo{journal}{{\em SN Computer Science\/}} \bibinfo{volume}{1},
  \bibinfo{number}{2} (\bibinfo{year}{2020}), \bibinfo{pages}{1--7}.
\newblock


\bibitem[\protect\citeauthoryear{Baharon, Shi, and Llewellyn-Jones}{Baharon
  et~al\mbox{.}}{2015}]%
        {baharon2015new}
\bibfield{author}{\bibinfo{person}{Mohd~Rizuan Baharon}, \bibinfo{person}{Qi
  Shi}, {and} \bibinfo{person}{David Llewellyn-Jones}.}
  \bibinfo{year}{2015}\natexlab{}.
\newblock \showarticletitle{A new lightweight homomorphic encryption scheme for
  mobile cloud computing}. In \bibinfo{booktitle}{{\em 2015 IEEE International
  Conference on Computer and Information Technology; Ubiquitous Computing and
  Communications; Dependable, Autonomic and Secure Computing; Pervasive
  Intelligence and Computing}}. IEEE, \bibinfo{pages}{618--625}.
\newblock


\bibitem[\protect\citeauthoryear{Chen, Wan, Celesti, Li, Abbas, and Zhang}{Chen
  et~al\mbox{.}}{2018}]%
        {chen2018edge}
\bibfield{author}{\bibinfo{person}{Baotong Chen}, \bibinfo{person}{Jiafu Wan},
  \bibinfo{person}{Antonio Celesti}, \bibinfo{person}{Di Li},
  \bibinfo{person}{Haider Abbas}, {and} \bibinfo{person}{Qin Zhang}.}
  \bibinfo{year}{2018}\natexlab{}.
\newblock \showarticletitle{Edge computing in IoT-based manufacturing}.
\newblock \bibinfo{journal}{{\em IEEE Communications Magazine\/}}
  \bibinfo{volume}{56}, \bibinfo{number}{9} (\bibinfo{year}{2018}),
  \bibinfo{pages}{103--109}.
\newblock


\bibitem[\protect\citeauthoryear{Gai, Wu, Zhu, Xu, and Zhang}{Gai
  et~al\mbox{.}}{2019}]%
        {gai2019permissioned}
\bibfield{author}{\bibinfo{person}{Keke Gai}, \bibinfo{person}{Yulu Wu},
  \bibinfo{person}{Liehuang Zhu}, \bibinfo{person}{Lei Xu}, {and}
  \bibinfo{person}{Yan Zhang}.} \bibinfo{year}{2019}\natexlab{}.
\newblock \showarticletitle{Permissioned blockchain and edge computing
  empowered privacy-preserving smart grid networks}.
\newblock \bibinfo{journal}{{\em IEEE Internet of Things Journal\/}}
  \bibinfo{volume}{6}, \bibinfo{number}{5} (\bibinfo{year}{2019}),
  \bibinfo{pages}{7992--8004}.
\newblock


\bibitem[\protect\citeauthoryear{Guo, Hu, Guo, Qiu, and Qi}{Guo
  et~al\mbox{.}}{2019}]%
        {guo2019blockchain}
\bibfield{author}{\bibinfo{person}{Shaoyong Guo}, \bibinfo{person}{Xing Hu},
  \bibinfo{person}{Song Guo}, \bibinfo{person}{Xuesong Qiu}, {and}
  \bibinfo{person}{Feng Qi}.} \bibinfo{year}{2019}\natexlab{}.
\newblock \showarticletitle{Blockchain meets edge computing: A distributed and
  trusted authentication system}.
\newblock \bibinfo{journal}{{\em IEEE Transactions on Industrial
  Informatics\/}} \bibinfo{volume}{16}, \bibinfo{number}{3}
  (\bibinfo{year}{2019}), \bibinfo{pages}{1972--1983}.
\newblock


\bibitem[\protect\citeauthoryear{Gupta}{Gupta}{2017}]%
        {gupta2017blockchain}
\bibfield{author}{\bibinfo{person}{Sourav~Sen Gupta}.}
  \bibinfo{year}{2017}\natexlab{}.
\newblock \showarticletitle{Blockchain}.
\newblock \bibinfo{journal}{{\em IBM Onlone (http://www. IBM. COM)\/}}
  (\bibinfo{year}{2017}).
\newblock


\bibitem[\protect\citeauthoryear{Hamdaoui, Alkalbani, Znati, and
  Rayes}{Hamdaoui et~al\mbox{.}}{2019}]%
        {hamdaoui2019unleashing}
\bibfield{author}{\bibinfo{person}{Bechir Hamdaoui}, \bibinfo{person}{Mohamed
  Alkalbani}, \bibinfo{person}{Taieb Znati}, {and} \bibinfo{person}{Ammar
  Rayes}.} \bibinfo{year}{2019}\natexlab{}.
\newblock \showarticletitle{Unleashing the power of participatory IoT with
  blockchains for increased safety and situation awareness of smart cities}.
\newblock \bibinfo{journal}{{\em IEEE Network\/}} \bibinfo{volume}{34},
  \bibinfo{number}{2} (\bibinfo{year}{2019}), \bibinfo{pages}{202--209}.
\newblock


\bibitem[\protect\citeauthoryear{Jangirala, Das, and Vasilakos}{Jangirala
  et~al\mbox{.}}{2019}]%
        {jangirala2019designing}
\bibfield{author}{\bibinfo{person}{Srinivas Jangirala},
  \bibinfo{person}{Ashok~Kumar Das}, {and} \bibinfo{person}{Athanasios~V
  Vasilakos}.} \bibinfo{year}{2019}\natexlab{}.
\newblock \showarticletitle{Designing secure lightweight blockchain-enabled
  RFID-based authentication protocol for supply chains in 5G mobile edge
  computing environment}.
\newblock \bibinfo{journal}{{\em IEEE Transactions on Industrial
  Informatics\/}} \bibinfo{volume}{16}, \bibinfo{number}{11}
  (\bibinfo{year}{2019}), \bibinfo{pages}{7081--7093}.
\newblock


\bibitem[\protect\citeauthoryear{Johnson, Menezes, and Vanstone}{Johnson
  et~al\mbox{.}}{2001}]%
        {johnson2001elliptic}
\bibfield{author}{\bibinfo{person}{Don Johnson}, \bibinfo{person}{Alfred
  Menezes}, {and} \bibinfo{person}{Scott Vanstone}.}
  \bibinfo{year}{2001}\natexlab{}.
\newblock \showarticletitle{The elliptic curve digital signature algorithm
  (ECDSA)}.
\newblock \bibinfo{journal}{{\em International journal of information
  security\/}} \bibinfo{volume}{1}, \bibinfo{number}{1} (\bibinfo{year}{2001}),
  \bibinfo{pages}{36--63}.
\newblock


\bibitem[\protect\citeauthoryear{Kang, Yu, Huang, Wu, Maharjan, Xie, and
  Zhang}{Kang et~al\mbox{.}}{2018}]%
        {kang2018blockchain}
\bibfield{author}{\bibinfo{person}{Jiawen Kang}, \bibinfo{person}{Rong Yu},
  \bibinfo{person}{Xumin Huang}, \bibinfo{person}{Maoqiang Wu},
  \bibinfo{person}{Sabita Maharjan}, \bibinfo{person}{Shengli Xie}, {and}
  \bibinfo{person}{Yan Zhang}.} \bibinfo{year}{2018}\natexlab{}.
\newblock \showarticletitle{Blockchain for secure and efficient data sharing in
  vehicular edge computing and networks}.
\newblock \bibinfo{journal}{{\em IEEE Internet of Things Journal\/}}
  \bibinfo{volume}{6}, \bibinfo{number}{3} (\bibinfo{year}{2018}),
  \bibinfo{pages}{4660--4670}.
\newblock


\bibitem[\protect\citeauthoryear{Krylovskiy}{Krylovskiy}{2015}]%
        {krylovskiy2015internet}
\bibfield{author}{\bibinfo{person}{Alexandr Krylovskiy}.}
  \bibinfo{year}{2015}\natexlab{}.
\newblock \showarticletitle{Internet of things gateways meet linux containers:
  Performance evaluation and discussion}. In \bibinfo{booktitle}{{\em 2015 IEEE
  2nd World Forum on Internet of Things (WF-IoT)}}. IEEE,
  \bibinfo{pages}{222--227}.
\newblock


\bibitem[\protect\citeauthoryear{Lab}{Lab}{}]%
        {nknwebsite}
\bibfield{author}{\bibinfo{person}{NKN Lab}.}
\newblock \bibinfo{title}{NKN: a Scalable Self-Evolving and Self-Incentivized
  Decentralized Network}.
\newblock   (\bibinfo{year}{????}).
\newblock
\showURL{%
\url{https://docs.nkn.org/docs/introduction-to-nkn}}


\bibitem[\protect\citeauthoryear{Liu, Ning, Xiong, and Yang}{Liu
  et~al\mbox{.}}{2014}]%
        {liu2014shared}
\bibfield{author}{\bibinfo{person}{Hong Liu}, \bibinfo{person}{Huansheng Ning},
  \bibinfo{person}{Qingxu Xiong}, {and} \bibinfo{person}{Laurence~T Yang}.}
  \bibinfo{year}{2014}\natexlab{}.
\newblock \showarticletitle{Shared authority based privacy-preserving
  authentication protocol in cloud computing}.
\newblock \bibinfo{journal}{{\em IEEE Transactions on parallel and distributed
  systems\/}} \bibinfo{volume}{26}, \bibinfo{number}{1} (\bibinfo{year}{2014}),
  \bibinfo{pages}{241--251}.
\newblock


\bibitem[\protect\citeauthoryear{Liu, Zhang, and Yang}{Liu
  et~al\mbox{.}}{2018}]%
        {liu2018blockchain}
\bibfield{author}{\bibinfo{person}{Hong Liu}, \bibinfo{person}{Yan Zhang},
  {and} \bibinfo{person}{Tao Yang}.} \bibinfo{year}{2018}\natexlab{}.
\newblock \showarticletitle{Blockchain-enabled security in electric vehicles
  cloud and edge computing}.
\newblock \bibinfo{journal}{{\em IEEE Network\/}} \bibinfo{volume}{32},
  \bibinfo{number}{3} (\bibinfo{year}{2018}), \bibinfo{pages}{78--83}.
\newblock


\bibitem[\protect\citeauthoryear{Mahajan and Sachdeva}{Mahajan and
  Sachdeva}{2013}]%
        {mahajan2013study}
\bibfield{author}{\bibinfo{person}{Prerna Mahajan} {and}
  \bibinfo{person}{Abhishek Sachdeva}.} \bibinfo{year}{2013}\natexlab{}.
\newblock \showarticletitle{A study of encryption algorithms AES, DES and RSA
  for security}.
\newblock \bibinfo{journal}{{\em Global Journal of Computer Science and
  Technology\/}} (\bibinfo{year}{2013}).
\newblock


\bibitem[\protect\citeauthoryear{Osvik, Bos, Stefan, and Canright}{Osvik
  et~al\mbox{.}}{2010}]%
        {osvik2010fast}
\bibfield{author}{\bibinfo{person}{Dag~Arne Osvik}, \bibinfo{person}{Joppe~W
  Bos}, \bibinfo{person}{Deian Stefan}, {and} \bibinfo{person}{David
  Canright}.} \bibinfo{year}{2010}\natexlab{}.
\newblock \showarticletitle{Fast software AES encryption}. In
  \bibinfo{booktitle}{{\em International Workshop on Fast Software
  Encryption}}. Springer, \bibinfo{pages}{75--93}.
\newblock


\bibitem[\protect\citeauthoryear{Rathore, Kwon, and Park}{Rathore
  et~al\mbox{.}}{2019}]%
        {rathore2019blockseciotnet}
\bibfield{author}{\bibinfo{person}{Shailendra Rathore},
  \bibinfo{person}{Byung~Wook Kwon}, {and} \bibinfo{person}{Jong~Hyuk Park}.}
  \bibinfo{year}{2019}\natexlab{}.
\newblock \showarticletitle{BlockSecIoTNet: Blockchain-based decentralized
  security architecture for IoT network}.
\newblock \bibinfo{journal}{{\em Journal of Network and Computer
  Applications\/}}  \bibinfo{volume}{143} (\bibinfo{year}{2019}),
  \bibinfo{pages}{167--177}.
\newblock


\bibitem[\protect\citeauthoryear{Ren, Zhao, Guan, and Lin}{Ren
  et~al\mbox{.}}{2020}]%
        {ren2020novel}
\bibfield{author}{\bibinfo{person}{Yan Ren}, \bibinfo{person}{Qiuxia Zhao},
  \bibinfo{person}{Haipeng Guan}, {and} \bibinfo{person}{Zhiqiang Lin}.}
  \bibinfo{year}{2020}\natexlab{}.
\newblock \showarticletitle{A novel authentication scheme based on edge
  computing for blockchain-based distributed energy trading system}.
\newblock \bibinfo{journal}{{\em EURASIP Journal on Wireless Communications and
  Networking\/}} \bibinfo{volume}{2020}, \bibinfo{number}{1}
  (\bibinfo{year}{2020}), \bibinfo{pages}{1--15}.
\newblock


\bibitem[\protect\citeauthoryear{Shi, Cao, Zhang, Li, and Xu}{Shi
  et~al\mbox{.}}{2016}]%
        {shi2016edge}
\bibfield{author}{\bibinfo{person}{Weisong Shi}, \bibinfo{person}{Jie Cao},
  \bibinfo{person}{Quan Zhang}, \bibinfo{person}{Youhuizi Li}, {and}
  \bibinfo{person}{Lanyu Xu}.} \bibinfo{year}{2016}\natexlab{}.
\newblock \showarticletitle{Edge computing: Vision and challenges}.
\newblock \bibinfo{journal}{{\em IEEE internet of things journal\/}}
  \bibinfo{volume}{3}, \bibinfo{number}{5} (\bibinfo{year}{2016}),
  \bibinfo{pages}{637--646}.
\newblock


\bibitem[\protect\citeauthoryear{Sitt{\'o}n-Candanedo, Alonso, Corchado,
  Rodr{\'\i}guez-Gonz{\'a}lez, and Casado-Vara}{Sitt{\'o}n-Candanedo
  et~al\mbox{.}}{2019}]%
        {sitton2019review}
\bibfield{author}{\bibinfo{person}{In{\'e}s Sitt{\'o}n-Candanedo},
  \bibinfo{person}{Ricardo~S Alonso}, \bibinfo{person}{Juan~M Corchado},
  \bibinfo{person}{Sara Rodr{\'\i}guez-Gonz{\'a}lez}, {and}
  \bibinfo{person}{Roberto Casado-Vara}.} \bibinfo{year}{2019}\natexlab{}.
\newblock \showarticletitle{A review of edge computing reference architectures
  and a new global edge proposal}.
\newblock \bibinfo{journal}{{\em Future Generation Computer Systems\/}}
  \bibinfo{volume}{99} (\bibinfo{year}{2019}), \bibinfo{pages}{278--294}.
\newblock


\bibitem[\protect\citeauthoryear{Song, Gu, Feng, Ge, and Mohapatra}{Song
  et~al\mbox{.}}{2021}]%
        {song2020blockchain}
\bibfield{author}{\bibinfo{person}{Jinyue Song}, \bibinfo{person}{Tianbo Gu},
  \bibinfo{person}{Xiaotao Feng}, \bibinfo{person}{Yunjie Ge}, {and}
  \bibinfo{person}{Prasant Mohapatra}.} \bibinfo{year}{2021}\natexlab{}.
\newblock \showarticletitle{Blockchain meets COVID-19: A framework for contact
  information sharing and risk notification system}. In
  \bibinfo{booktitle}{{\em 2021 IEEE 17th International Conference on Mobile Ad
  Hoc and Sensor Systems (MASS)}}. IEEE.
\newblock


\bibitem[\protect\citeauthoryear{Song, Gu, Ge, and Mohapatra}{Song
  et~al\mbox{.}}{2020}]%
        {song2020smart}
\bibfield{author}{\bibinfo{person}{Jinyue Song}, \bibinfo{person}{Tianbo Gu},
  \bibinfo{person}{Yunjie Ge}, {and} \bibinfo{person}{Prasant Mohapatra}.}
  \bibinfo{year}{2020}\natexlab{}.
\newblock \showarticletitle{Smart Contract-based Computing Resources Trading in
  Edge Computing}. In \bibinfo{booktitle}{{\em 2020 IEEE 31st Annual
  International Symposium on Personal, Indoor and Mobile Radio
  Communications}}. IEEE, \bibinfo{pages}{1--7}.
\newblock


\bibitem[\protect\citeauthoryear{Usman, Jan, Jolfaei, Xu, He, and Chen}{Usman
  et~al\mbox{.}}{2019}]%
        {usman2019distributed}
\bibfield{author}{\bibinfo{person}{Muhammad Usman}, \bibinfo{person}{Mian~Ahmad
  Jan}, \bibinfo{person}{Alireza Jolfaei}, \bibinfo{person}{Min Xu},
  \bibinfo{person}{Xiangjian He}, {and} \bibinfo{person}{Jinjun Chen}.}
  \bibinfo{year}{2019}\natexlab{}.
\newblock \showarticletitle{A distributed and anonymous data collection
  framework based on multilevel edge computing architecture}.
\newblock \bibinfo{journal}{{\em IEEE Transactions on Industrial
  Informatics\/}} \bibinfo{volume}{16}, \bibinfo{number}{9}
  (\bibinfo{year}{2019}), \bibinfo{pages}{6114--6123}.
\newblock


\bibitem[\protect\citeauthoryear{Wang, Feng, Chen, George, Bala, Pillai, Yang,
  and Satyanarayanan}{Wang et~al\mbox{.}}{2018}]%
        {wang2018bandwidth}
\bibfield{author}{\bibinfo{person}{Junjue Wang}, \bibinfo{person}{Ziqiang
  Feng}, \bibinfo{person}{Zhuo Chen}, \bibinfo{person}{Shilpa George},
  \bibinfo{person}{Mihir Bala}, \bibinfo{person}{Padmanabhan Pillai},
  \bibinfo{person}{Shao-Wen Yang}, {and} \bibinfo{person}{Mahadev
  Satyanarayanan}.} \bibinfo{year}{2018}\natexlab{}.
\newblock \showarticletitle{Bandwidth-efficient live video analytics for drones
  via edge computing}. In \bibinfo{booktitle}{{\em 2018 IEEE/ACM Symposium on
  Edge Computing (SEC)}}. IEEE, \bibinfo{pages}{159--173}.
\newblock


\bibitem[\protect\citeauthoryear{Wang, Wu, Choo, and He}{Wang
  et~al\mbox{.}}{2019}]%
        {wang2019blockchain}
\bibfield{author}{\bibinfo{person}{Jing Wang}, \bibinfo{person}{Libing Wu},
  \bibinfo{person}{Kim-Kwang~Raymond Choo}, {and} \bibinfo{person}{Debiao He}.}
  \bibinfo{year}{2019}\natexlab{}.
\newblock \showarticletitle{Blockchain-based anonymous authentication with key
  management for smart grid edge computing infrastructure}.
\newblock \bibinfo{journal}{{\em IEEE Transactions on Industrial
  Informatics\/}} \bibinfo{volume}{16}, \bibinfo{number}{3}
  (\bibinfo{year}{2019}), \bibinfo{pages}{1984--1992}.
\newblock


\bibitem[\protect\citeauthoryear{Wang, Ji, Guo, Yu, Chen, and Li}{Wang
  et~al\mbox{.}}{2020}]%
        {wang2020trafficchain}
\bibfield{author}{\bibinfo{person}{Qianlong Wang}, \bibinfo{person}{Tianxi Ji},
  \bibinfo{person}{Yifan Guo}, \bibinfo{person}{Lixing Yu},
  \bibinfo{person}{Xuhui Chen}, {and} \bibinfo{person}{Pan Li}.}
  \bibinfo{year}{2020}\natexlab{}.
\newblock \showarticletitle{TrafficChain: A blockchain-based secure and
  privacy-preserving traffic map}.
\newblock \bibinfo{journal}{{\em IEEE Access\/}}  \bibinfo{volume}{8}
  (\bibinfo{year}{2020}), \bibinfo{pages}{60598--60612}.
\newblock


\bibitem[\protect\citeauthoryear{Xu, Jiang, Wang, Yuan, and Ren}{Xu
  et~al\mbox{.}}{2014}]%
        {xu2014information}
\bibfield{author}{\bibinfo{person}{Lei Xu}, \bibinfo{person}{Chunxiao Jiang},
  \bibinfo{person}{Jian Wang}, \bibinfo{person}{Jian Yuan}, {and}
  \bibinfo{person}{Yong Ren}.} \bibinfo{year}{2014}\natexlab{}.
\newblock \showarticletitle{Information security in big data: privacy and data
  mining}.
\newblock \bibinfo{journal}{{\em Ieee Access\/}}  \bibinfo{volume}{2}
  (\bibinfo{year}{2014}), \bibinfo{pages}{1149--1176}.
\newblock


\bibitem[\protect\citeauthoryear{Xu, Zhang, Gao, Xue, Qi, and Dou}{Xu
  et~al\mbox{.}}{2019}]%
        {xu2019become}
\bibfield{author}{\bibinfo{person}{Xiaolong Xu}, \bibinfo{person}{Xuyun Zhang},
  \bibinfo{person}{Honghao Gao}, \bibinfo{person}{Yuan Xue},
  \bibinfo{person}{Lianyong Qi}, {and} \bibinfo{person}{Wanchun Dou}.}
  \bibinfo{year}{2019}\natexlab{}.
\newblock \showarticletitle{BeCome: Blockchain-enabled computation offloading
  for IoT in mobile edge computing}.
\newblock \bibinfo{journal}{{\em IEEE Transactions on Industrial
  Informatics\/}} \bibinfo{volume}{16}, \bibinfo{number}{6}
  (\bibinfo{year}{2019}), \bibinfo{pages}{4187--4195}.
\newblock


\bibitem[\protect\citeauthoryear{Zhang, Chen, Zhao, Cheng, and Hu}{Zhang
  et~al\mbox{.}}{2018}]%
        {zhang2018data}
\bibfield{author}{\bibinfo{person}{Jiale Zhang}, \bibinfo{person}{Bing Chen},
  \bibinfo{person}{Yanchao Zhao}, \bibinfo{person}{Xiang Cheng}, {and}
  \bibinfo{person}{Feng Hu}.} \bibinfo{year}{2018}\natexlab{}.
\newblock \showarticletitle{Data security and privacy-preserving in edge
  computing paradigm: Survey and open issues}.
\newblock \bibinfo{journal}{{\em IEEE access\/}}  \bibinfo{volume}{6}
  (\bibinfo{year}{2018}), \bibinfo{pages}{18209--18237}.
\newblock


\bibitem[\protect\citeauthoryear{Zhang, Mao, Leng, He, and Zhang}{Zhang
  et~al\mbox{.}}{2017}]%
        {zhang2017mobile}
\bibfield{author}{\bibinfo{person}{Ke Zhang}, \bibinfo{person}{Yuming Mao},
  \bibinfo{person}{Supeng Leng}, \bibinfo{person}{Yejun He}, {and}
  \bibinfo{person}{Yan Zhang}.} \bibinfo{year}{2017}\natexlab{}.
\newblock \showarticletitle{Mobile-edge computing for vehicular networks: A
  promising network paradigm with predictive off-loading}.
\newblock \bibinfo{journal}{{\em IEEE Vehicular Technology Magazine\/}}
  \bibinfo{volume}{12}, \bibinfo{number}{2} (\bibinfo{year}{2017}),
  \bibinfo{pages}{36--44}.
\newblock


\bibitem[\protect\citeauthoryear{Zhang, Hong, Chen, Zheng, and Chen}{Zhang
  et~al\mbox{.}}{2019}]%
        {zhang2019joint}
\bibfield{author}{\bibinfo{person}{Zhen Zhang}, \bibinfo{person}{Zicong Hong},
  \bibinfo{person}{Wuhui Chen}, \bibinfo{person}{Zibin Zheng}, {and}
  \bibinfo{person}{Xu Chen}.} \bibinfo{year}{2019}\natexlab{}.
\newblock \showarticletitle{Joint computation offloading and coin loaning for
  blockchain-empowered mobile-edge computing}.
\newblock \bibinfo{journal}{{\em IEEE Internet of Things Journal\/}}
  \bibinfo{volume}{6}, \bibinfo{number}{6} (\bibinfo{year}{2019}),
  \bibinfo{pages}{9934--9950}.
\newblock


\bibitem[\protect\citeauthoryear{Zhao, Hwang, and Villeta}{Zhao
  et~al\mbox{.}}{2012}]%
        {zhao2012game}
\bibfield{author}{\bibinfo{person}{Zhou Zhao}, \bibinfo{person}{Kai Hwang},
  {and} \bibinfo{person}{Jose Villeta}.} \bibinfo{year}{2012}\natexlab{}.
\newblock \showarticletitle{Game cloud design with virtualized CPU/GPU servers
  and initial performance results}. In \bibinfo{booktitle}{{\em Proceedings of
  the 3rd workshop on Scientific Cloud Computing}}. \bibinfo{pages}{23--30}.
\newblock


\bibitem[\protect\citeauthoryear{Zhou, Shi, Ouyang, and Zhao}{Zhou
  et~al\mbox{.}}{2021}]%
        {zhou2021building}
\bibfield{author}{\bibinfo{person}{Huan Zhou}, \bibinfo{person}{Zeshun Shi},
  \bibinfo{person}{Xue Ouyang}, {and} \bibinfo{person}{Zhiming Zhao}.}
  \bibinfo{year}{2021}\natexlab{}.
\newblock \showarticletitle{Building a blockchain-based decentralized ecosystem
  for cloud and edge computing: an ALLSTAR approach and empirical study}.
\newblock \bibinfo{journal}{{\em Peer-to-Peer Networking and Applications\/}}
  (\bibinfo{year}{2021}), \bibinfo{pages}{1--17}.
\newblock


\end{thebibliography}
